\documentclass[twocolumn,amsmath,amssymb,aps,showpacs,superscriptaddress]{revtex4}

\usepackage{graphicx}
\usepackage{amsmath}
\usepackage{amssymb}
\usepackage{xcolor}
\usepackage{dcolumn}
\usepackage{hyperref}

\begin{document}

\title{Analysis of short term price trends in daily stock-market
index data}

\author{H.F. Coronel-Brizio}
\affiliation{ Departamento de Inteligencia Artificial, Facultad de F\'{\i}sica e Inteligencia Artificial, Universidad Veracruzana. Sebastian Camacho 5, Xalapa, Veracruz 91000, M\'exico}
\author{A.R. Hern\'andez Montoya}
\email{alhernandez@uv.mx}
\affiliation{ Departamento de Inteligencia Artificial, Facultad de F\'{\i}sica e Inteligencia Artificial, Universidad Veracruzana. Sebastian Camacho 5, Xalapa, Veracruz 91000, M\'exico}
\author{H.R Olivares S\'anchez}
\email{rolivares@fis.cinvestav.mx}
\affiliation{Departamento de F\'{\i}sica, CINVESTAV-IPN, Av. IPN 2508, Col. San Pedro Zacatenco, M\'exico D.F., Mexico}
\affiliation{Departamento de F\'{\i}sica, Facultad de F\'{\i}sica e Inteligencia Artificial, Universidad Veracruzana. Circuito Gonzalo Aguirre Beltr\'an s/n, Zona Universitaria, Xalapa, Veracruz 91000, M\'exico}
\author{Enrico Scalas}
\email{enrico.scalas@unipmn.it}
\homepage{http://people.unipmn.it/scalas/}
\affiliation{Dipartimento di Scienze e Innovazione Tecnologica,
Universit\`a del Piemonte Orientale ``Amedeo Avogadro'', Viale T. Michel 11, 15121 Alessandria, Italy}
\affiliation{BCAM-Basque Center for Applied Mathematics, Alameda de Mazarredo 14, 48009 Bilbao, Basque Contry, Spain}

\date{\today}

\begin{abstract}
In financial time series there are periods in which the value increases or decreases monotonically. We call those periods {\it elemental trends} and study the probability distribution of their duration for the indices DJIA, NASDAQ and IPC. It is found that the trend duration distribution often differs from the one expected under no memory. The expected and observed distributions are compared by means of the Anderson-Darling test.
\end{abstract}


\maketitle

\section{Introduction}
\label{intro}
One of the goals of financial-market analysis is to predict the future movements of prices and financial indices. In order to achieve this goal, a huge variety of methods to forecast markets behavior were developed, ranging from complex mathematical models even to astrological pseudo-scientific techniques. An approach that has been recently growing in popularity is the statistical analysis of large sets of data, which has become now possible due to the increasing availability of computer power and high quality data sets. This approach has benefited from the contributions not only from economists, but also from many physicists and mathematicians who have applied methods and ideas of probability theory and statistical physics to finance. A set of nontrivial statistical properties of historical data was observed and classified as ``stylized facts'' \cite{Rama}, which are expected to provide a better insight on market structure and behavior.

When observing the time series of the prices of an asset on a chart, it is common to see ``trends'' in which most of the values are greater (or smaller) than the previous ones. These trends are very popular within the so-called {\it technical analysis}. Trends as those studied by technical analysis can be seen as composed by smaller elemental trends, periods in which the value increases or decreases monotonically. These kind of trends are the ones that will be studied in the present work. Among other things, technical analysts seek patterns in the charts of financial data, that are believed to be indicators of changes in the trend direction. The effectiveness of technical analysis is disputed and put at a stake by what is known as the Efficient Market Hypothesis (EMH). Before going further, it is necessary to give some definitions. In Subsections \ref{def}, \ref{efficient} and \ref{stylized} of this introduction these definitions and other useful information will be presented. In Section \ref{model}, a model for the distribution of trends durations will be developed from the EMH. Section \ref{methodology} will explain how the data were analyzed and section \ref{discussion}  will provide an interpretation of the analysis.
\subsection{Definitions}
\label{def}
Let $S(t)$ be the price of an asset or an index value at time $t$ and $X(t) = \log{S(t)}$ its logarithm. The log-return at time $t$ is defined as:
\begin{equation}
r(t,\Delta{t}) = X(t + \Delta{t}) - X(t)
\end{equation}
for a given time sampling scale $\Delta{t}$. If the price variation is small, the log-return is a good approximation of the return
\begin{equation}
R(t,\Delta t) = \frac{S(t+\Delta t) - S(t)}{S(t)}.
\end{equation}
In this paper, we consider  $\Delta{t}$ equal to 1
trading day and we use the daily close values of the indices to
build the series $S(t)$. More details on the data sets will be given in section \ref{methodology}.

\begin{figure}
\begin{center}
\includegraphics[width=\columnwidth]{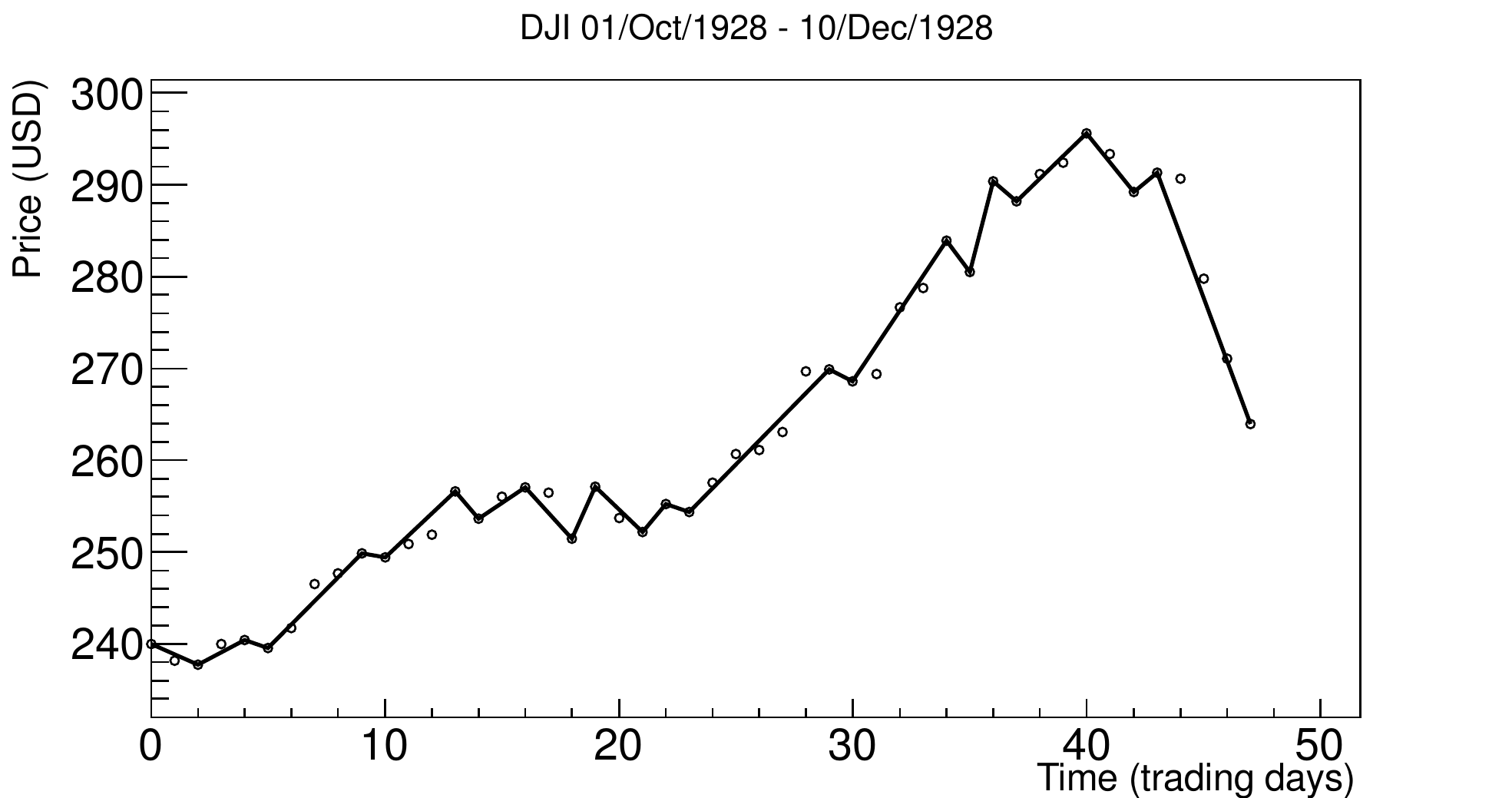}
\end{center}
\caption{\label{fig:trends}The line segments join the starting and ending points of each elemental trend.}
\end{figure}

An {\it elemental trend} of duration $k$ will be defined here as a subseries of $k+1$ values within the series $S(t)$ in which every value is greater (for an uptrend) or smaller or equal (for a downtrend) than the preceding one (Figure \ref{fig:trends}). The aim of this work is to study with a statistical approach the kind of short term trends defined above.

\subsection{The Efficient Market Hypothesis}
\label{efficient}
The EMH claims that the market quickly finds the rational price for a traded asset \cite{Mantegna}. The most important consequence of this hypothesis was shown by P. Samuelson \cite{Samuelson 2} and it is the fact that the best forecast for the future price of an asset is its present price.
\begin{equation}
\label{emh1}
\mathbb{E}(S(t+\Delta t)|\mathcal{F}_t ) = S(t),
\end{equation} 
where $\mathbb{E}(\cdot|\mathcal{F}_t)$ is the conditional expectation with respect to the filtration $\mathcal{F}_t$, namely with respect to the known history up to time $t$. Indeed, it is easy to derive the EMH from a simple no-arbitrage argument. Suppose we have two assets, a risky one, with price $S(t)$ and a risk-free one giving a constant interest rate $r_F$. To avoid arbitrage, one has to require that the expected return of the risky asset is equal to the risk-free interes rate, that is
\begin{equation}
\mathbb{E} (R(t, \Delta t)|\mathcal{F}_t) = r_F;
\end{equation}
the latter equation immediately yields, for non vanishing $S(t)$,
\begin{equation}
\label{emh2}
\mathbb{E}(S(t+\Delta t)|\mathcal{F}_t ) = (1+r_F) S(t),
\end{equation}
which reduces to \eqref{emh1} for $r_F = 0$. Equations \eqref{emh1} and
\eqref{emh2}, jointly with the integrability of the process $S(t)$, are known as martingale and sub-martingale (remember that $r_F \geq 0$) conditions, respectively.

The EMH would invalidate the attempts of technical analysis to predict future prices or trends; in fact, in Samuelson's words, ``there is no way of making an expected profit by extrapolating past changes in the futures price, by chart or any esoteric devices of magic or mathematics'' \cite{Samuelson 2} as the best forecast of the future price would be the current price.

\subsection{Stylized facts}
\label{stylized}
As mentioned before, financial time series share some nontrivial statistical properties called stylized facts. Although those properties are often formulated qualitatively, they are so constraining that it is difficult to reproduce all of them by means of a stochastic process \cite{Rama}. As a matter of fact, none of the market models, including analytical models, Monte Carlo simulations and multi-agent based models, created before 1990, when awareness of such regularities gradually started to appear, could reproduce all of these stylized facts \cite{Lux}. As an interesting issue, some studies suggest that stylized facts appear not only in financial time series, but also in other complex systems such as Conway's Game of Life \cite{Hernandez}. To fix the ideas, some of the stylized facts, taken from reference \cite{Rama}, are listed below:

\begin{description}
\item[Absence of linear autocorrelations:] Autocorrelations of returns are often negligible, except for very small time scales, depending on the market and on the time horizon.
\item[Heavy tails:] The return distribution is leptokurtic and some authors claim that the tails decay as a power-law.
\item[Gain-loss asymmetry:] Large downward jumps in stock prices and stock index values are observed, but not equally large upward movements. (In exchange rates there is a higher symmetry in up/down movements).
\item[Volatility clustering:] High volatility events do cluster in time.
\end{description}

\section{An `Efficient Market' model for the duration distribution}
\label{model}
Among all the possible martingale or sub-martingale models that can describe price fluctuations, the geometric random walk is the simplest one. A geometric random walk is just a product of independent and identically distributed positive random variables. If the expected value of these variables is $1$, then the geometric random walk is a martingale; otherwise, if the expected value is larger than $1$, the geometric random walk is a submartingale. However, the geometric random walk hypothesis is neither necessary nor sufficient for an efficient market, as shown by many authors among whom Leroy \cite{leroy}, Lucas \cite{lucas} and Lo and Mckinlay \cite{lo}. To understand this point, it is enough to consider Equation \eqref{emh2} allowing for any martingale model.

At each step of a series of index values, there are two possible outcomes: the index either increases or does not increase. In an efficient market, the expected future price depends only on information about the current price, not on its previous history. Therefore, it should be impossible to predict the expected direction of a future price change given the history of the price process. In formula, from Equation \eqref{emh1} (after discounting for the risk-free rate), we have\begin{equation}
\label{expectedincrement}
\mathbb{E}(S(t+\Delta t) - S(t)|\mathcal{F}_t)= 0;
\end{equation}
if we consider the sign of the price change $Y(t,\Delta t)=\mathrm{sign}(S(t+\Delta t)-S(t))$, which coincides with the sign of returns, we accordingly have
\begin{equation}
\label{expectedsign}
\mathbb{E}(Y(t,\Delta t))=0.
\end{equation}

If the price follows a geometric random walk, then the series of price-change signs can be modeled as a Bernoulli process. This process could be biased to take the presence of a risk free interest rate into account. To be more specific, let us consider a log-normal geometric random walk and let us use the assumption $\Delta t =1$. Let $S_0$ be the initial price. The price at time $t$ will be given by
\begin{equation}
S(t)=S_0 \prod_{i=1}^t Q_i
\end{equation}
where $Q_i$ are independent and identically distributed random variables following a log-normal distribution with parameters $\mu$ and $\sigma$. These two parameters come from the corresponding normal distribution for log-returns. As a direct consequence of the EMH in the form \eqref{emh2}, we have
\begin{equation}
\mathbb{E} (Q) = 1 + r_F,
\end{equation}
and for a log-normal distributed random variable, we have also
\begin{equation}
\mathbb{E}(Q) = \mathrm{e}^\mu \mathrm{e}^{\sigma^2/2}.
\end{equation}
This leads to a dependence between the two parameters
\begin{equation}
\mu = \log(1+r_F) - \frac{\sigma^2}{2}.
\end{equation}
Note that, when $r_F = 0$, it is impossible to get $\mu=0$. This reflects a more general result, if the price process is a martingale, the log-price process cannot be a martingale and viceversa. Starting from the cumulative distribution function for a log-normal random variable
\begin{equation}
F_Q(u)=\mathbb{P}(Q \leq u) = \frac{1}{2} + \frac{1}{2} \mathrm{erf}\left(
\frac{\log(u) -\mu}{\sqrt{2 \sigma^2}} \right),
\end{equation}
the probability of a negative sign would be given by
\begin{equation}
q=F_Q(1)=\mathbb{P}(Q \leq 1)=\frac{1}{2} + \frac{1}{2} \mathrm{erf}\left(
\frac{\sigma}{2 \sqrt{2}} - \frac{\log (1+r_F)}{\sigma \sqrt{2}} \right),
\end{equation}
which yields $q=1/2$ for $r_F = \mathrm{e}^{\sigma^2/2}-1$.

It becomes natural to use the biased Bernoulli process as the null hypothesis for the time series of signs \cite{scalasold}. It is well known that the distribution of the number $k$ of failures needed to get one success for a Bernoulli process with success probability $p=1-q$ is the geometric distribution ${\cal{G}}(p)$; the number of failures $N$ is given by
\begin{equation}
P(k)=\mathbb{P}(N=k) = p(1 - p)^{k} = pq^{k}.
\end{equation}
The duration of an elemental downward trend is the number of days before the price increases, so the distribution of such trend durations should follow a geometric distribution. An identical argument applies to the duration of an upward trend. Such sequences of identical outcomes are also known as {\em runs} or {\em clumps} in the mathematical literature.

\section{Methodology}
\label{methodology}
\subsection{Data}
Three indices were analyzed, namely Dow Jones Industrial Average (DJIA), NASDAQ Composite and the Mexican \'Indice de Precios y Cotizaciones (IPC) during the periods 1 October 1928 - 7 July 2011, 5 February 1971 - 30 September 2011 and 30 October 1978 - 7 April 2011, respectively. The data were taken from Yahoo Finance.

\subsection{Price and time scales}
The prices were expressed in terms of ``constant money'' using the consumer price indices from references \cite{cpi} and \cite{inpc}. Since the values of those indices were delivered monthly, linear interpolation was used in order to express the daily values. The time was measured in trading days as discussed above.

\subsection{Building the sample}
\textcolor{black}{For each series of index values, several time windows of 1000 trading
days were created, each one shifted forward by one trading day with respect to the
previous one. This procedure resulted in 20785 time windows for the DJIA, 10282
for the NASDAQ and 4997 for the IPC. Two histograms were built for
each time window, one for upward trends and the other one for
downward trends. For upward trends, 500 points within each time window
were selected at random, and the number of days before the first time decrease
were measured. Such waiting times were the entries for each
histogram. For instance, given the sequence $--+++++-$, assume that the fourth entry is randomly 
chosen. Then the recorded waiting time is 4.
The histograms for downward trends were built in the same
way. Finally, all the histograms were normalized. Examples
of both uptrend and downtrend histograms for different time windows
are shown in Figure \ref{tabone}.}

\subsection{The Anderson-Darling goodness of fit test}
In order to compare the observed and expected distributions of  trend durations, the Anderson-Darling test described in references \cite{Anderson,Choulakian} was used. The Anderson-Darling test was found to be the most suitable for this purpose because it places more weight on the tails of a distribution than other goodness of fit tests. The critical values of the Anderson-Darling statistic $A^{2}_{n}$ were dependent on the parameter of the geometric distribution, and they were estimated using Monte Carlo simulations.

\section{Discussion}
\label{discussion}

\begin{figure}
\begin{center}
\includegraphics[width=\columnwidth]{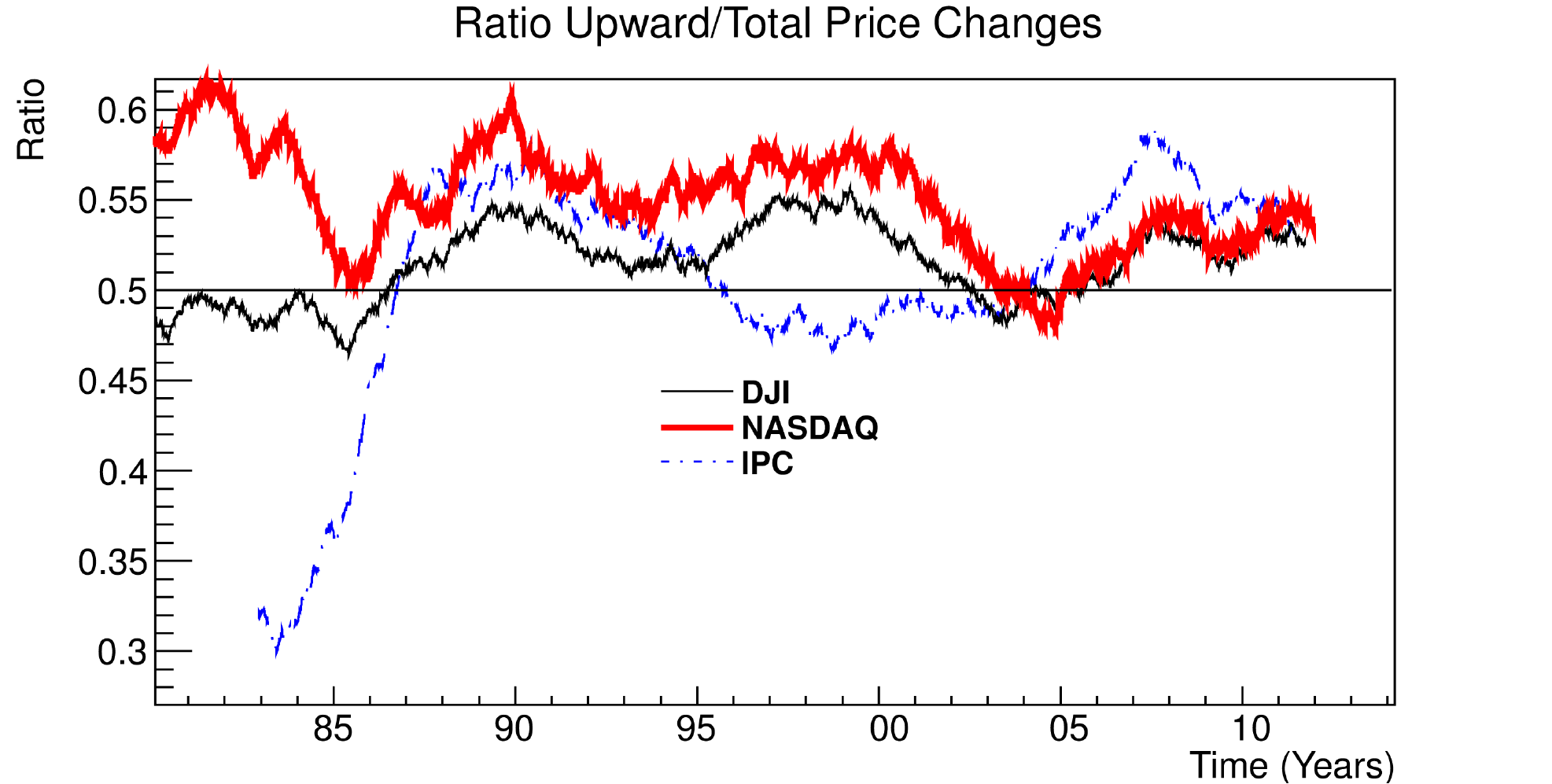}
\end{center}
\caption{\label{fig:simetria} (Color online) Ratio of upward to total price changes in daily data, plotted against time for the years 1980 - 2011, calculated over a time window of 1000 trading days.}
\end{figure}

\begin{figure*}[htbp]
\includegraphics[width=\columnwidth]{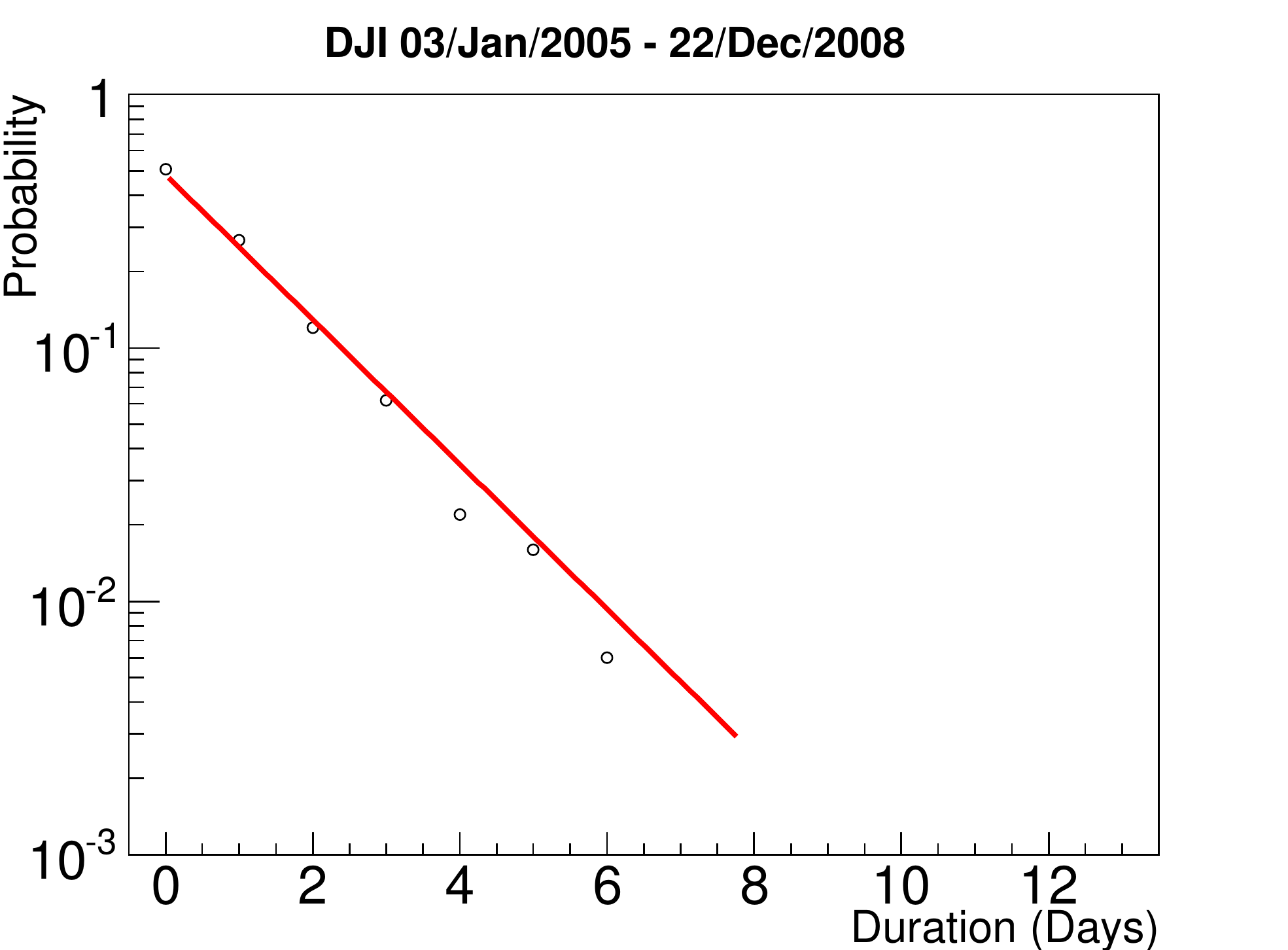}
\includegraphics[width=\columnwidth]{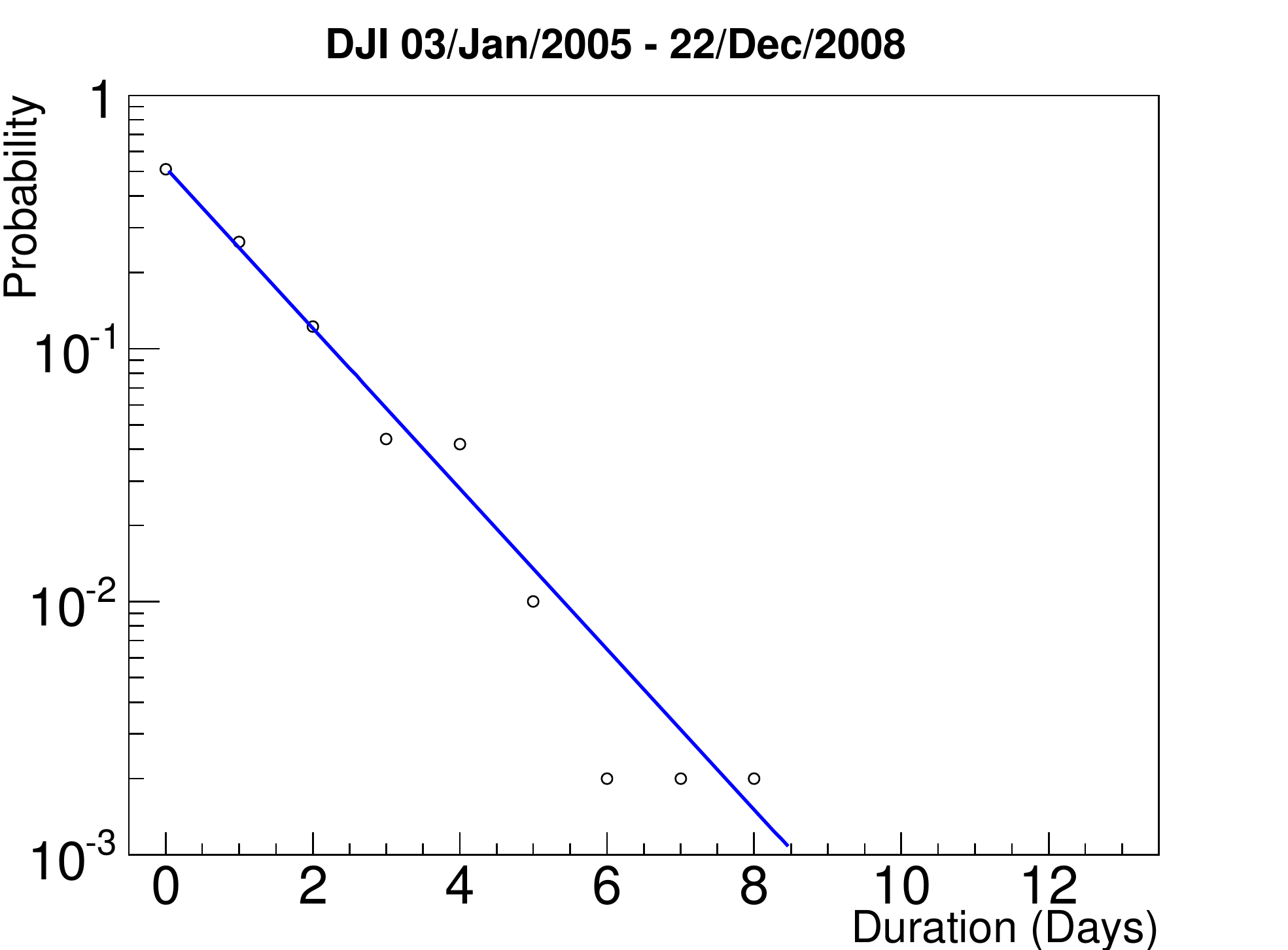}
\includegraphics[width=\columnwidth]{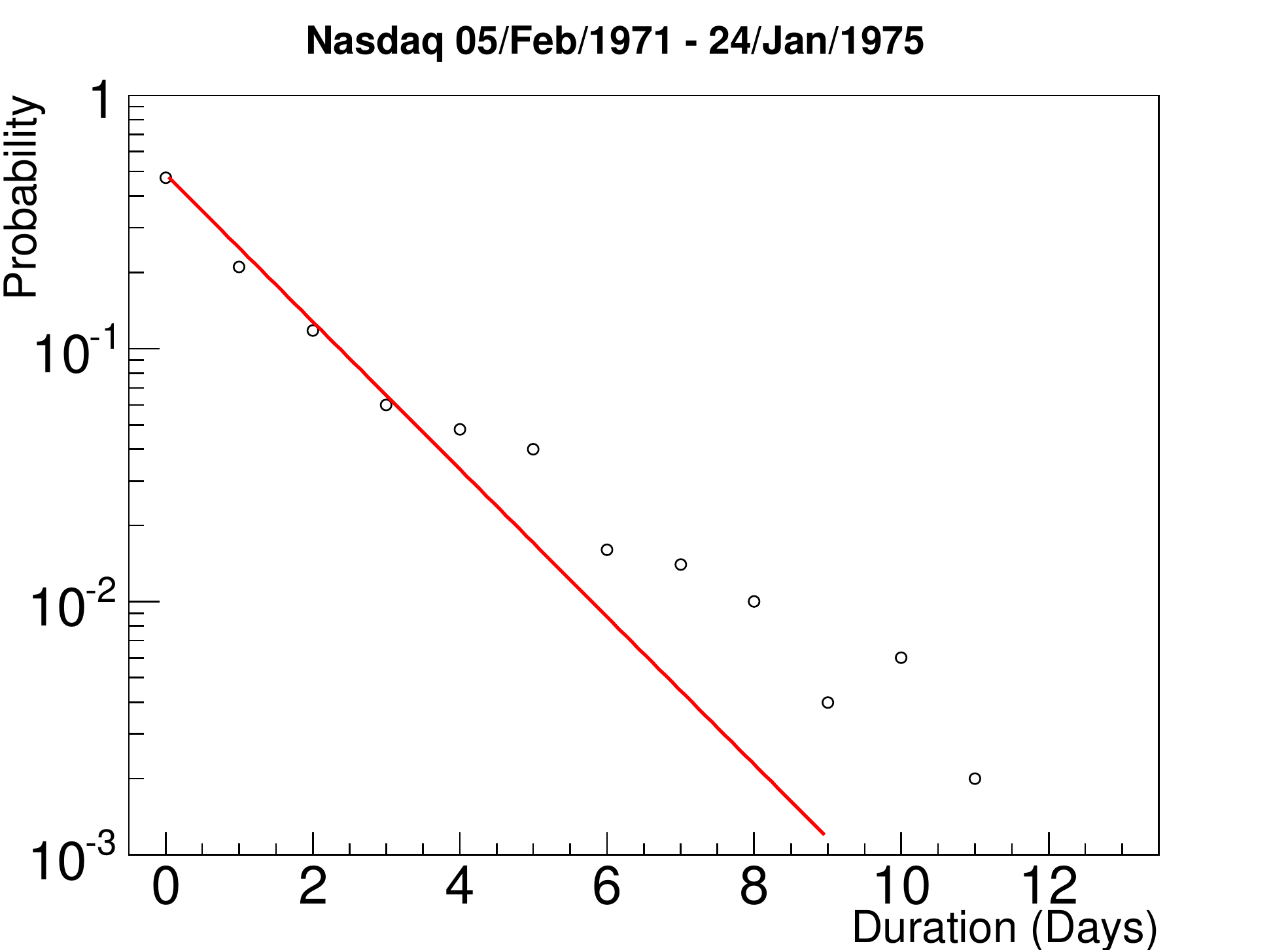}
\includegraphics[width=\columnwidth]{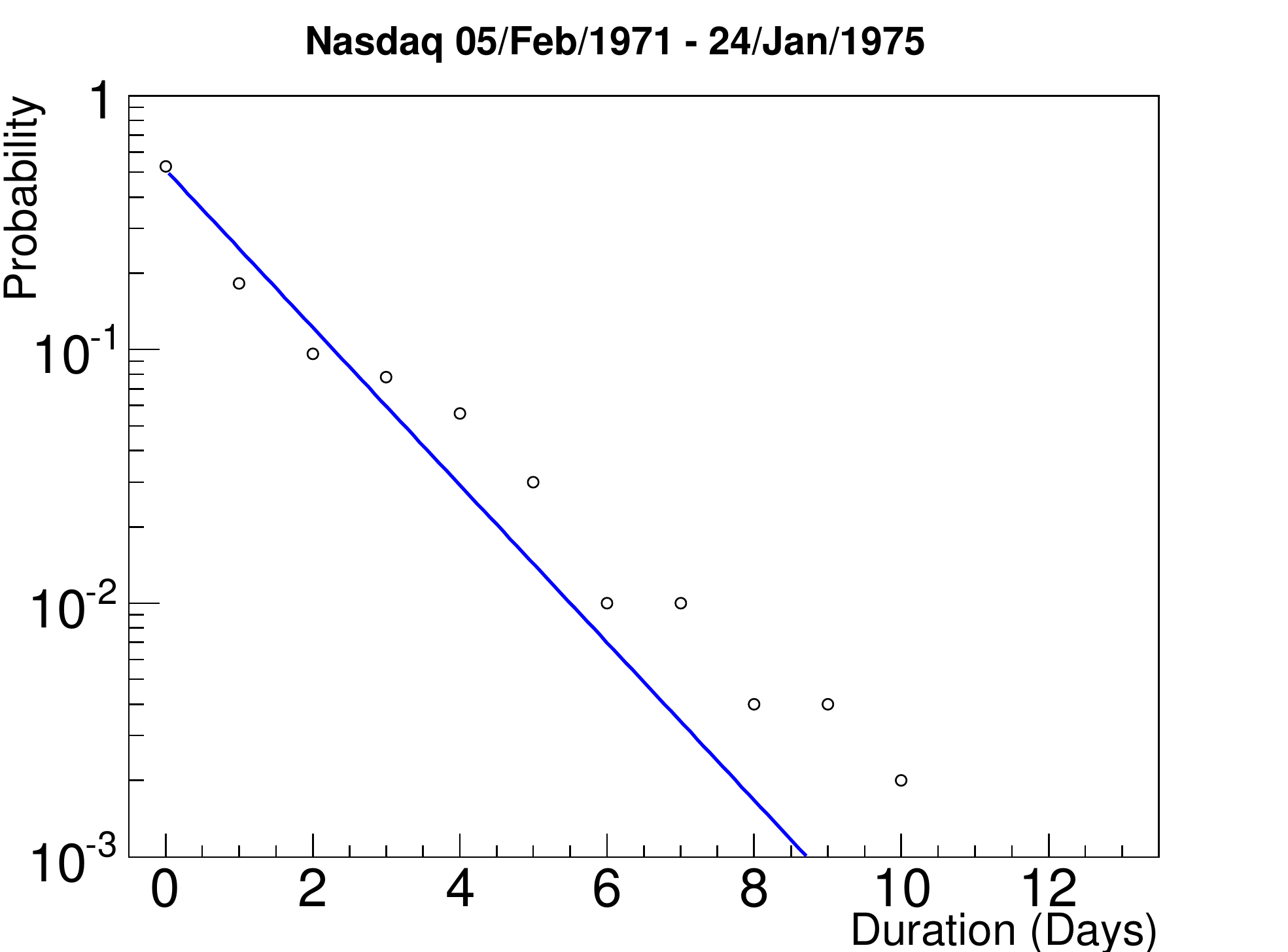}
\includegraphics[width=\columnwidth]{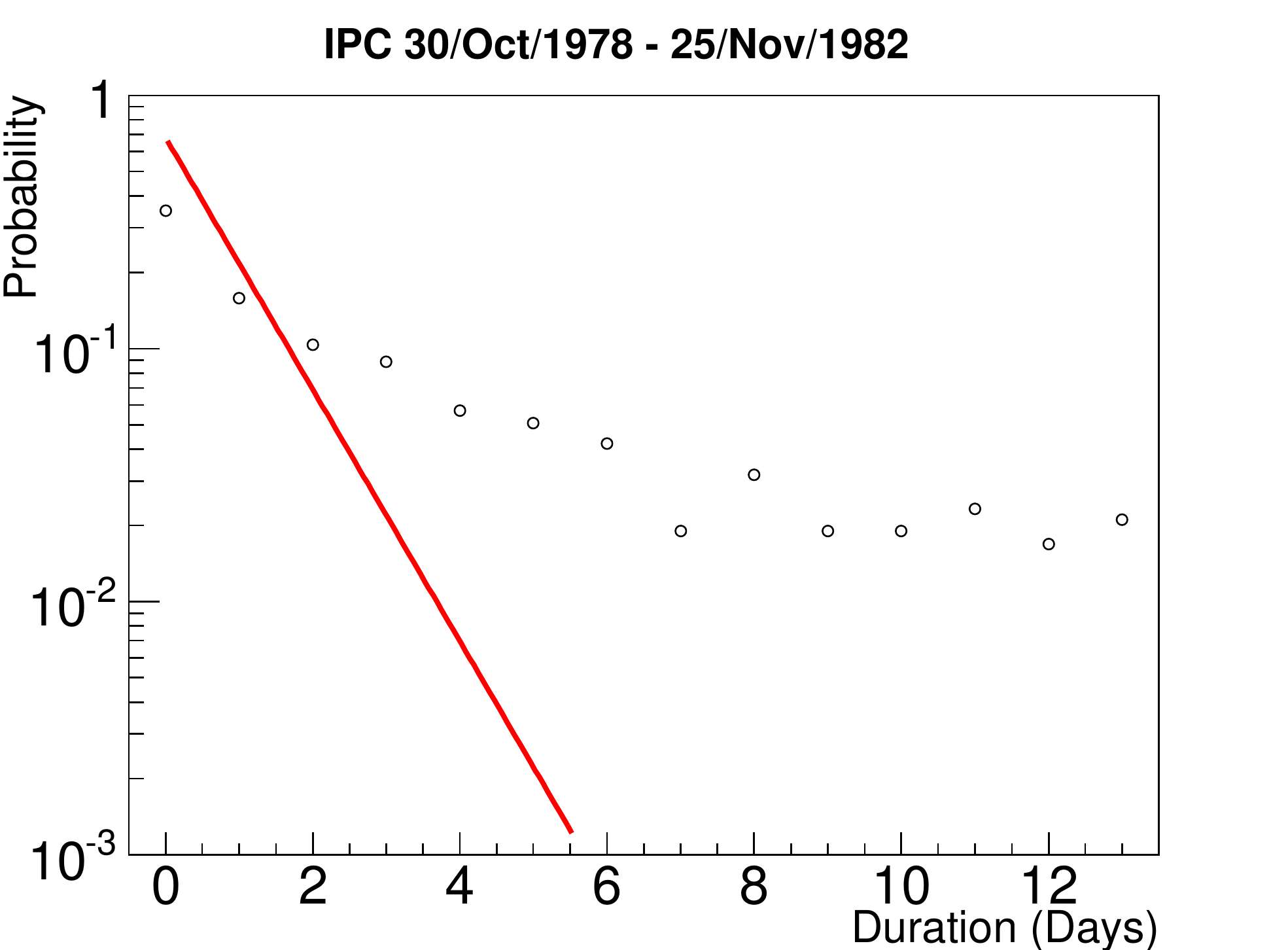}
\includegraphics[width=\columnwidth]{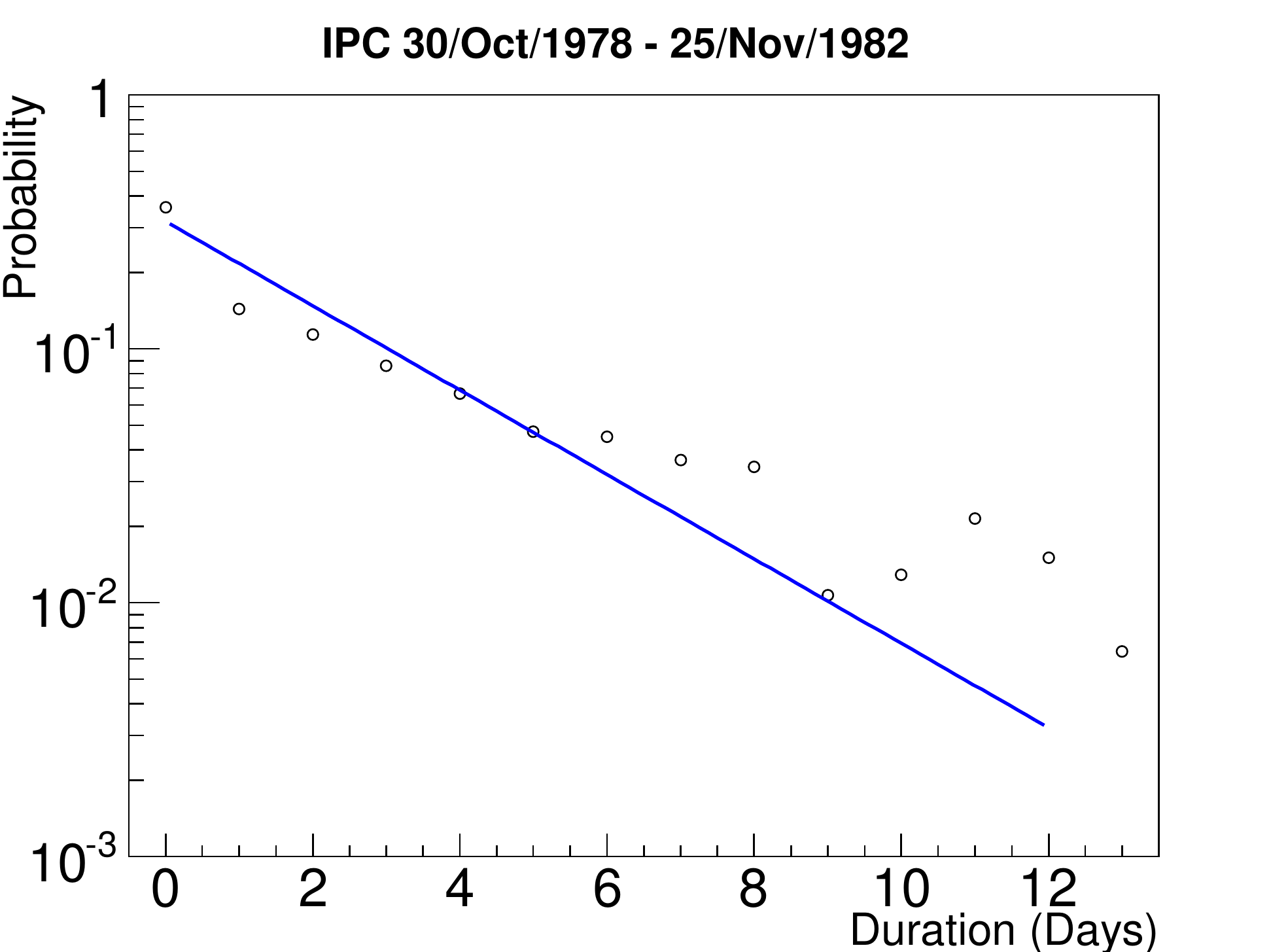}
\caption{\label{tabone} (Color online)
Distribution of trend durations for different indices and time windows of 1000 days. The red geometric fits on the left concern the upward distributions and the blue geometric fits on the right are performed on the downward empirical distributions. The estimate $\hat{p}$ of the parameter $p$ is given by 
the ratio between upward price changes and total price changes, whereas $\hat{q} = 1-\hat{p}$.
The estimated parameters $\hat{p}$ and $\hat{q}$ are as follow: first line left $\hat{p} = 0.518$ and first line right $\hat{q} = 0.482$,
second line left $\hat{p} = 0.511$ and second line right $\hat{q} = 0.489$, third line left  $\hat{p} = 0.318$ and third line right $\hat{q} = 0.682$.}
\end{figure*}

\begin{figure*}[htbp]
\includegraphics[width=\columnwidth]{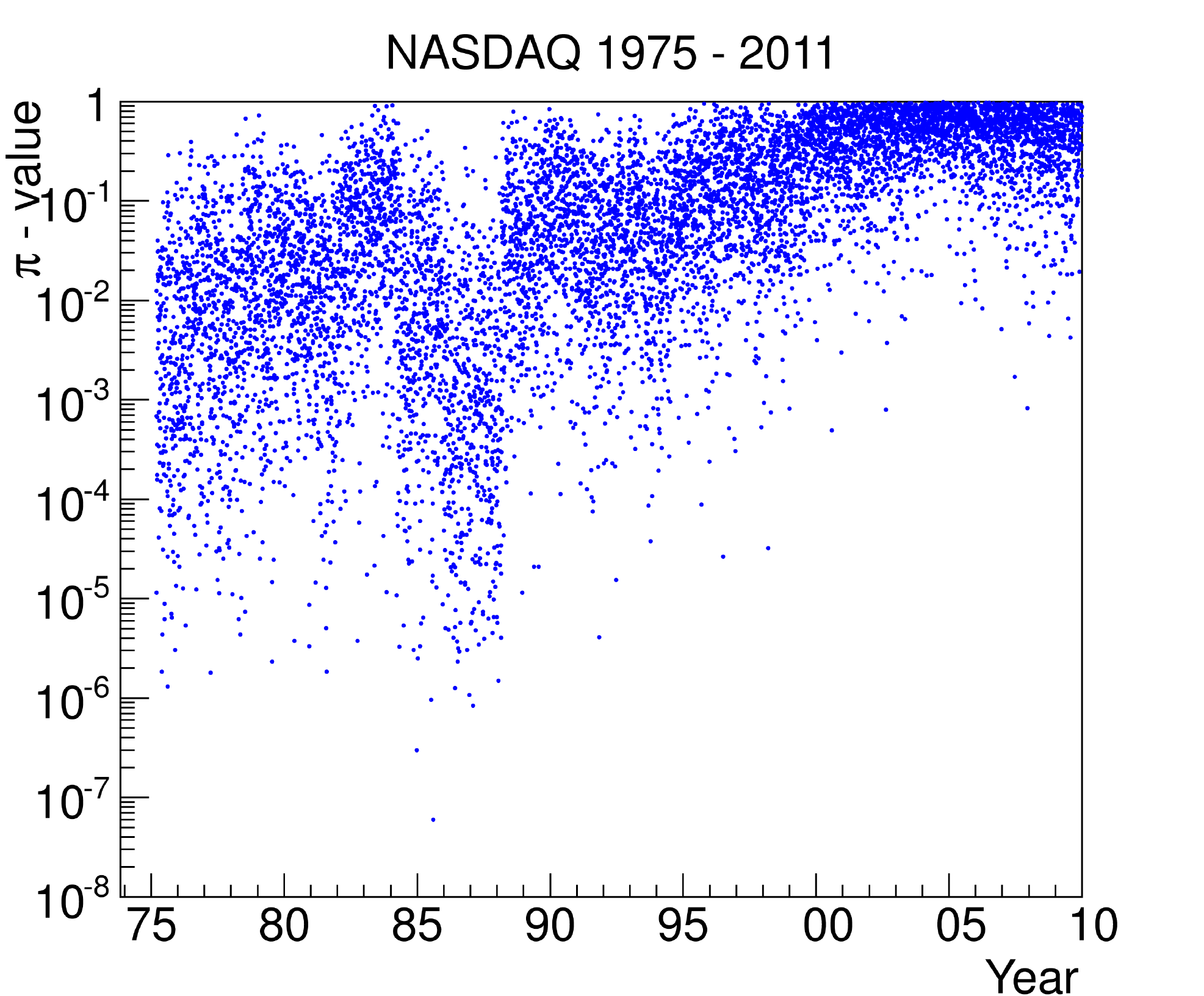}
\includegraphics[width=\columnwidth]{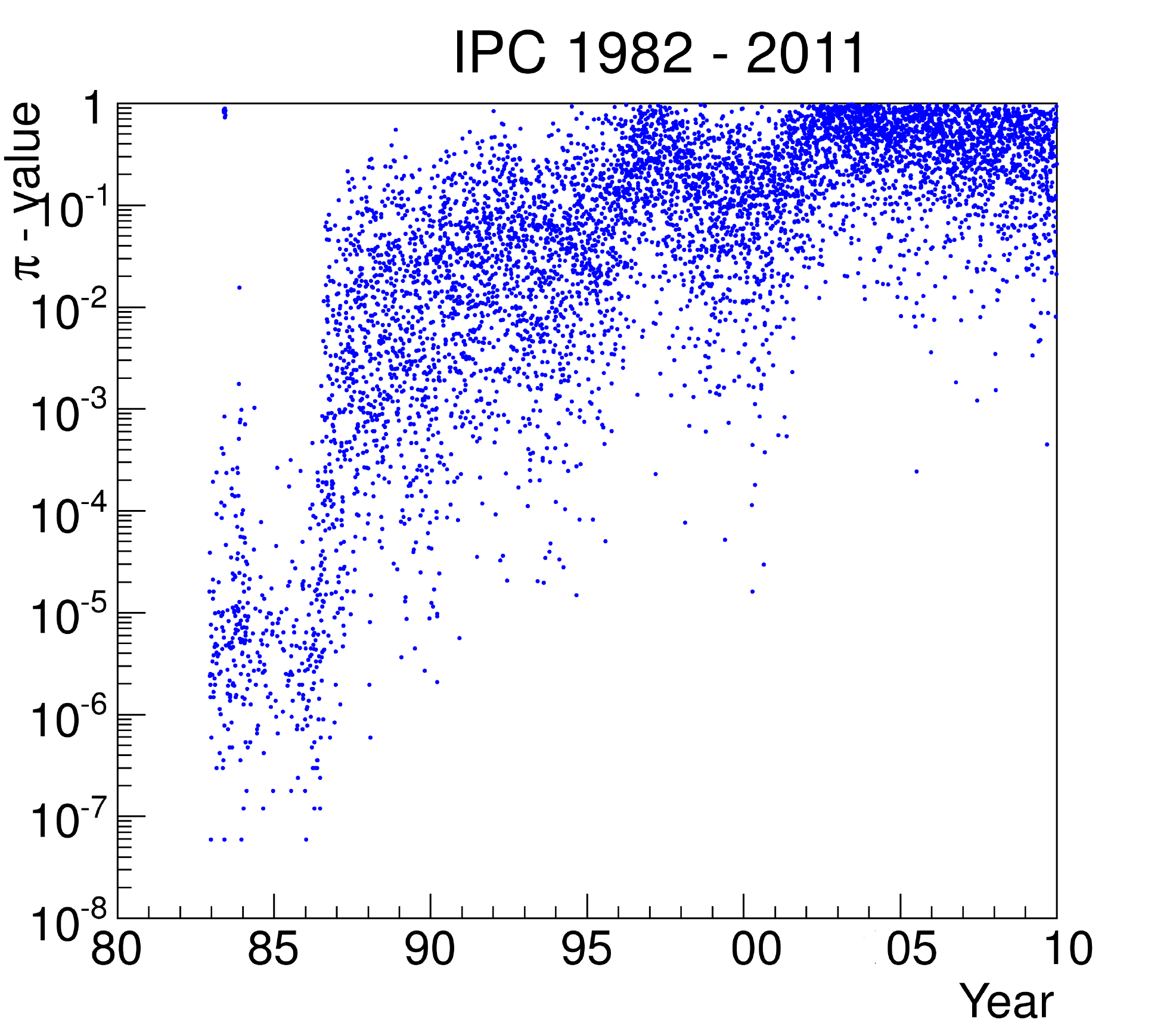}
\includegraphics[width=\columnwidth]{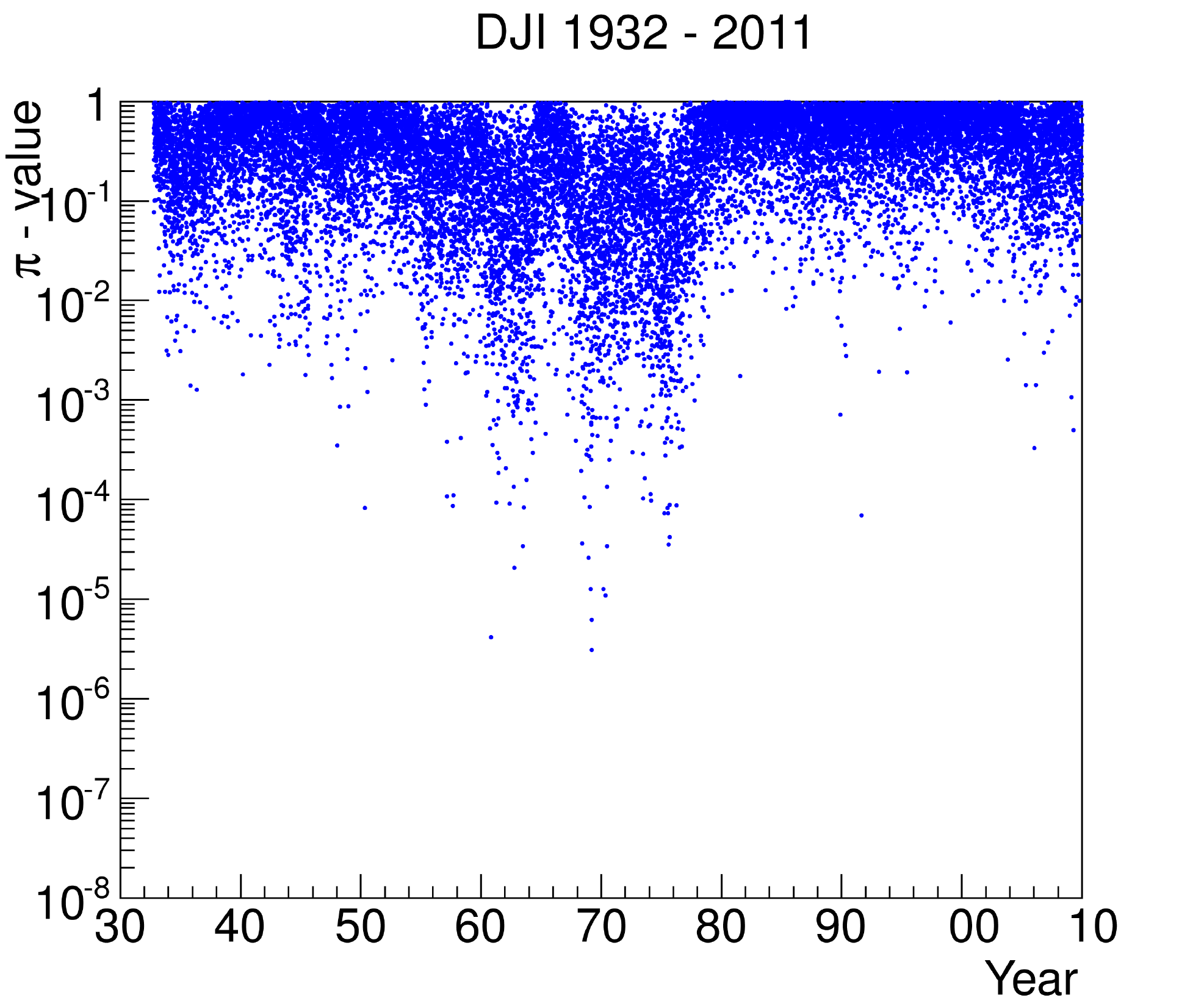}
\caption{\label{fig:nasdaqjpcdji} (Color online) Top left: $\pi$-values of the Anderson-Darling statistic for the NASDAQ plotted against time. As time passes, the data agree better with the geometric distribution; top right: $\pi$-values of the Anderson-Darling statistic for the IPC plotted against time. As for NASDAQ, agreement between data and the geometric distribution increases with time; bottom: $\pi$-values of the Anderson-Darling statistic for the DJIA plotted against time. The greatest deviations from the geometric distribution occurred between the years 1960-1980.}
\end{figure*}

In Figure \ref{fig:simetria}, the ratio of the upward to total price changes in daily data is plotted against time for the years 1980 - 2011. This ratio is calculated over a time window of 1000 trading days. It can be seen that variations are greater than those expected for the same time windows in a Bernoulli process with parameter $p = 1/2$ ($\pm{0.05}$), but it might be interesting to find out whether the hypothesis of a geometric distribution holds for smaller periods (such as each of the 1000 days time windows individually), because it would mean that in those periods the direction of price changes was not predictable using historical prices of the index. Figure \ref{tabone} show the distribution of trend durations corresponding to different indices and periods of 1000 days. Figure \ref{fig:nasdaqjpcdji} displays the $p$-values of the Anderson-Darling statistic for different periods. In order to avoid confusion between the parameter of the geometric distribution and the $p$-values for the distribution of $A^{2}_{n}$, the latter will be referred to as $\pi$-values. The meaning of $\pi$-values is the probability of obtaining a value of $A^{2}_{n}$ at least as big as the one that was really obtained, given that the probability distribution is actually geometric.

It was observed that as time passes, the direction of price changes for the IPC and the Nasdaq is better described by a geometric distribution (Figure \ref{fig:nasdaqjpcdji}). The distribution of trend durations for the Dow Jones is generally reasonably well fitted by the geometric distribution. This fact can be interpreted as a possible evidence that the Mexican stock market (that has become public and regulated since 1975) has been increasing its efficiency, as reported by previous research \cite{Achach}. The same claim can be made about the NASDAQ, given that it is also a market of relatively recent creation. In contrast, the Dow Jones Industrial Average index represents a more mature market. However, there is also evidence that the New York Stock Exchange, represented by the Dow Jones, has swiftly increased its efficiency between the beginning of the 1980s and the end of the 1990s \cite{NYSE}. Figure \ref{fig:nasdaqjpcdji} shows that for the Dow Jones, the greatest deviations from the geometric distribution in the studied period (almost the whole XXth Century) occurred between the years 1960-1980.

\section{Conclusions}

The probability distributions for the duration of elemental trends were studied for the market indices Dow Jones Industrial Average (DJIA), NASDAQ Composite and for the Mexican \'Indice de Precios y Cotizaciones (IPC). These distributions are expected to be geometric and memoryless according to the discussion in section \ref{model}. The IPC and the NASDAQ present periods in which the memoryless hypothesis must be definitely rejected.

\section*{Acknowledgements}
This work was supported by Conacyt-Mexico and MAE-Italy under grant 146498. We also thank Conacyt-Mexico for the support under  project grant 155492, Universidad Veracruzana under project 41504 and PRIN 2009 Italian grant ``Finitary and non-finitary probabilistic methods in Economics''. The IPC daily data set  from 1978 to 2006 was provided by P. Zorrilla-Velasco, A. Reynoso del Valle and S. Herrera-Montiel,  from  BMV at that time. We are grateful to all of them.


\begin{thebibliography}{15}

\bibitem{Anderson}
Bracquemond C, Cr\'etois E and Gaudoin O 2002 `A comparative study of goodness-of-fit tests for the geometric distribution and application to discrete time reliability' {\it Preprint} {\tt http://www-ljk.imag.fr/SMS/preprints.html}.

\bibitem{Choulakian}
V Choulakian, R A Lockhart, M A Stephens 1994 `Cramer-von Mises Statistics for Discrete Distributions'
Canadian Journal Of Statistics {\bf 22} 125--137


\bibitem{Rama}
Cont R 2001 `Empirical properties of asset returns: stylized facts and statistical issues' {\it Quantitative Finance} {\bf 1} 223--236

\bibitem{Achach}
Coronel-Brizio H F, Hern\'andez-Montoya A R, Huerta-Quintanilla R, and Rodr\'{\i}guez-Achach M E 2007 `Evidence of increment of efficiency of the Mexican Stock Market through the analysis of its variations', \textit{Physica A} (380) 391--398

\bibitem{Hernandez}
Hern\'andez Montoya A R, Coronel-Brizio H F, Rodr\'{\i}guez-Achach M E, Stevens-Ram\'{\i}rez G A, Politi M, and Scalas E 2011 `Emerging properties of financial time series in the Game of Life' {\it Phys. Rev.} E {\bf 84} 066104

\bibitem{leroy}
Leroy S F 1973, `Risk aversion and the martingale property of stock prices' {\it International Economic Review} {\bf 14} 436--446.

\bibitem{lucas} Lucas R E 1978, `Asset prices in an exchange economy'  {\it Econometrica} {\bf 46} 1429--1445.

\bibitem{lo} 
Lo A W and Mackinlay A C 1999, {\it A Non-Random Walk Down Wall Street}, (Princeton: Princeton University Press).

\bibitem{Holyst}
Ho\l{yst} J A and Sieczka P 2008 `Statistical properties of short term price trends in high frequency stock market data' {\it Physica} A {\bf 387} 1218--1224

\bibitem{inpc}
Instituto Nacional de Estad\'istica, Geograf\'ia e Inform\'atica. \'Indice Nacional de Precios al Consumidor. (Retrieved on 2011, 5 November). 
{\tt http://www.inegi.org.mx/est/contenidos/\\
proyectos/inp/inpc.aspx}.

\bibitem{gainloss}
Jensen M H, Johansen A and Simonsen I 2002 `Inverse Statistics in Economic: The gain-loss asymmetry', {\it Physica} A {\bf 324} 6.

\bibitem{Levy-Solomon}
Levy M, Levy H and Solomon S 2000 {\it Microsimulations of Financial Markets: From Investor Behavior to Market Phenomena} (London (UK): Academic Press)

\bibitem{Mantegna}
Mantegna R N and Stanley H E 2000 {\it An Introduction to Econophysics}. (Cambridge (UK): Cambridge University Press)

\bibitem{Lux}
Samanidou E, Zschischang E, Stauffer D and Lux T 2006 `Microscopic Models of Financial Markets'. Economics Working Paper 2006-15. Department of Economics, University of Kiel.

\bibitem{Samuelson 2}
Samuelson P A 1965 â `Proof that Properly Anticipated Prices Fluctuate Randomly', {\it Industrial Management Rev.} {\bf 6} 41--45.

\bibitem{Samuelson}
Samuelson P A and Nordhaus W D 2005 {\it Economics} (New York: McGraw-Hill).

\bibitem{scalasold} Scalas E 1998, `Scaling in the market of futures', {\it Physica} A {\bf 253} 394--402.

\bibitem{NYSE}
Toth B and Kertesz J `Increasing market efficiency: Evolution of cross-correlations of stock returns' {\it Physica} A {\bf 320} 505--515

\bibitem{cpi}
U. S. Bureau of Labor Statistics, Consumer Price Index. (Retrieved on 2011, 5 November). {\tt http://data.bls.gov/timeseries/ \\
CUSR0000SA0?output\_view=pct\_1mth}.

\bibitem{Voit}
Voit J 2003 {\it The Statistical Mechanics of Financial Markets}. (New York: Springer-Verlag).

\end{thebibliography}
\end{document}